# RADIATION IN A CLOSED 3-D UNIVERSE REVEALS ITS PRESENT GEOMETRY AND ITS PAST EVOLUTION

## Charles B. Leffert

Emeritus Professor, Wayne State University, Detroit, MI 48202
(c_leffert@wayne.edu)

**Abstract** Predictions of the new "Spatial Condensation (SC-)" cosmological model were presented on the foundations [1], the new source of gravity [2], and large-scale structure [3]. In this paper predictions will include new physical features that are due mostly to the postulated geometry of our closed 3-D universe. Knowledge of the past universe is obtained from its many types of radiation that travel to our instruments on great circles of our expanding 3-sphere. Adding the fourth spatial dimension greatly increases our understanding of the concepts of emission and reception distances of the sources and 4-D trajectory of the radiation. In a closed expanding 3-D universe, most of the radiation from a distant bright source at $Z_e$ can travel halfway around the universe and be refocused as a "virtual source" at $Z_v \sim 1.5$ where some of the re-diverging radiation arrives later to our instruments. With the added fourth dimension, the SC-model makes some spectacular predictions for these virtual sources. Indeed, the CBR becomes a re-focused "backside" microwave radiation with its structure amplified in size and decreased in luminosity by a factor $\sim 1/350$. "Viewed" in the optical from a direction opposite the CBR, the same, but more evolved, structure at "virtual $Z_{same}$" would certainly fix the geometry and size of our universe. Virtual sources are also added radiation to the diffuse background of X-rays and infrared. The CBR dipole suggests a preferred reference frame. Reflected light signals in a non-accelerated frame show, in principle, the moving frame could be brought to rest in the comoving frame.

## OVERVIEW

We live in a 3-D cellular space consisting of tiny Planck size spatial cells in violent agitation. This 3-D space is the surface of a 4-D ball that is expanding in the mother epi-space. Nothing has yet been postulated about a small 4-D thickness in the orthogonal direction of the radius of the 4-D core or whether there is more than one layer in that surface. To date, spatial condensation has needed only the postulate of the production of 4-D spatial cells to define energy and to expand the 4-D core and our 3-D universe. Radiation travels in that 3-D vacuum surface at the speed of light but how and in which layer is for future development.

From the expansion rate of our 3-D universe at the end of creation, it was concluded that the corresponding energy transport rate in the epi-universe is $C^+ > 10^{24} C$ and that enormous rate may account for the non-locality and some other mysteries of quantum behavior. [9]

In paper 3 [3] it was shown that at the present time, every volume of 3-D space reproduces its volume in 4.75 Gy and not only accounts for Hubble's law but also solves the "cosmological constant" problem. [4] At large radii spatial condensation produces an expansion force on any massive object that is prevented (such as by gravity) from participating freely in that Hubble flow.

The most important assumptions for this study of radiation are that the mother epi-universe is Euclidean and that the radiation travels on great circles of the expanding 4-D ball from emission to the astronomer's detector.

It was shown in paper $2^2$ that the SC-model of gravity predicts the same local curvature of 3-D space, as does general relativity. Therefore the gravitational lensing by large massive objects in the SC-model should be the same as in relativity theory; such local perturbations are not discussed.

As radiation is transferred from one spatial planckton to the next every Planck time at $C=l_p/t_p$, new 3-D planckton are injected into the photon to increase its wavelength, decrease its frequency and energy, $(dE_p/dt)/E_p=-H$ and thus account for its redshift. For an observer at rest in the comoving frame of an expanding SC-universe, the radial "compounded velocity $v_c$" of radiation at distance r is $v_c=Hr\pm C$.[9] Harrison[6] uses the outgoing +C version to define the recession velocity $U_p$ of the particle horizon ($r=L_p$) or $U_p=H_0L_p+C$, but interest here is in an incoming packet of radiation at $-C$:

$$v_c=Hr-C. \qquad (1)$$

The velocity $v_H=Hr$ is not a velocity in a medium (ether) in our 3-D space, it is a velocity of a point in space relative to a distant comoving point (e.g. in 3-D, the increasing distance between two points on an expanding balloon). Locally it is very small and cannot be measured with present instruments. The present SC-$H_0=2.2 \times 10^{-18}$ s$^{-1}$ gives only $v_H=3.6 \times 10^{-13}$ cm/s at r=one million miles.

Because of the way time is defined in the SC-computer model, the trajectory of a packet of radiation in 4-D space as our universe expands can be directly calculated using Eq. (1). Graphic displays of a number of such trajectories will be presented. Much of the qualitative analysis follows from the beginning postulate that our 3-D universe is a closed expanding 3-sphere.

## EXPANSION FUNCTIONS FOR RADIATION

For the radiation reaching an astronomer's instruments, the distance to the source at the time of emission or "emission distance, ED" and the distance to that source from the detector at reception or "reception distance, RD" are both dependent on the astronomer's model of the 3-D universe. Indeed, it might be thought that both distance concepts appear somewhat "fuzzy" because for early emission sources, neither a detector nor an observer existed for a "distance" to be "measured from" and at reception of the radiation, the source may have long ceased to exist. Nevertheless, both concepts are well defined because on the scale of the universe, both the atoms that will form the detector and the atoms of the source have well-defined world lines.

In the big-bang model, the Friedmann equation involves scale factor R, total mass-energy density $\rho$ and the curvature of space k. In turn, the Friedmann equation yields the comoving coordinate $r_1$ of the source whose radiation exhibits a redshift Z on detection at the present where the integral over r from 0 to r of $(1-kr^2)^{-1/2}$ gives a dimensionless distance $r_1R_0/(C/H_0)$ in terms of Z and $\Omega_0$ which is said to be independent of k when the pressure p of the universe is insignificant ($p=\omega\rho$ and $\omega=0$).[5]

For comparison to predictions from the SC-model, the solutions for ED and RD for the simple relativistic Einstein-de Sitter universe (n=2/3) in units of $L_H=C/H_0$ are:[6]

$$RD=2(1 - 1/(1+Z)^{1/2}), \qquad (2)$$
$$ED=(2/(1+Z))(1-1/(1+Z)^{1/2}), \qquad (3)$$

for this flat k=0 universe. But note that integration of Eq. (1) also yields Eqs. (2) and (3) for the Einstein-de Sitter universe [9] where $v_c$=C only for the comoving frame (r=0).
.
      A different approach will be taken for the SC-model. A packet of radiation from the position of emission ED will be followed through both 3-D and 4-D space as the universe expands until it arrives at the detector. Important to the interpretation is that each point along that trajectory also represents another point of emission for its radiation to reach the detector at the same time of detection. Although each source will be assumed to be at rest in 3-D space (peculiar velocity $v_p$=0) its distance at reception is simply RD=$(R_0/R_e)$ED=(1+Z)ED. The results and comparison of these two models are shown in Fig. 4-1.

      Note that the emission distances for both the Einstein-de Sitter (ES) universe and the spatial-condensation (SC) universe increase to a higher redshift Z~1.5 and then decrease. The derivative with respect to Z of Eq. (3-3) of Paper 3 set to zero,[3] has its maximum at $Z_m$=1.25 for the Einstein-de Sitter in good agreement with Fig. 4-1. For the SC-model, the maximum is somewhat higher at $Z_m$=1.70. To explain the maximum, first remember that the direction of travel of the packet of radiation is from high Z to Z=0 at the present. At distances of high Z, the expansion rate of space (relative to the detector) $v_H$==Hr is greater than the speed of light C toward the detector. Not until the maximum $Z_m$ does Hr-C = 0 and only after, as H continues to decrease with the expansion, does the compounded velocity $v_c$=Hr-C become negative and allow the wave packet to decrease the much larger distance to the detector. Finally, when r=0, $v_c$=-C into the detector.

      This behavior appears much more dramatic in Fig. 4-2 for the SC-model with the wave packet of radiation emitted at a much earlier redshift of 999 at decoupling of radiation and matter when our universe was only 1/1000 of its present size. In the early development of the model, the age in this computer run was set to 15 Gy. At the time the radiation was emitted, it was only ~67 light years (ly) from the future position of the detector. For the next 5 billion years, the rapid expansion carried the radiation packet away, against its forward speed of 307 Mpc/Gy to a distance of 5.72 billion ly and then it took another 10 billion years, still against the expansion, for the radiation packet to cover the growing 5.72 Gly.

      This apparent competition between the expansion velocity $v_H$=Hr and the speed of light C is made more clear in the 4-D space of the SC-model as shown in Fig. 4-3 for the same case as that of Figs. 4-1 and 4-2 [Note that limited scaling distorts circles in this figure.]. The first 2-D plot of the 3-D surface of the 4-D core using two of the 4-D coordinates was presented in Paper 2 [2] for the derived SC-profile of a non-rotating black hole. A few descriptive comments on such plots will be repeated here for the reader. Both X and Y are 4-D coordinates and thus only one 3-D coordinate can be shown and only as points on an expanding circle. If our 3-D space was flat, as preferred by present cosmologists, its edge-on view could well be represented by just the horizontal line Y=0 (infinitely long) in Fig. 4-3.

      However, the SC-model began with the assumption that our 3-D spatial universe was closed -- that it was the surface of an expanding 4-D ball – a topology not unlike many of the objects we see inside our 3-D universe from rain drops to radiant stars. Therefore, at any one cosmic time, our 3-D universe appears as a circle in Fig. 4-3. It would appear as a filled circle (cross-hatched) if the 4-D core was also represented.

Instead, the expansion of the 4-D ball is indicated by four successive circles of increasing radius R and circumference 2πR.

The largest circle represents our present universe, Z=0. The smallest circle containing the source of the cosmic background radiation (CBR), Z=999, when our universe was only 1/(1+Z) = 1/1000 of its present size, appears only as the center dot.

Start with assumption (A-1) that radiation travels on great circles on the surface of the expanding 4-D core. Our detector is at the top, call it a "telescope", and is pointed clockwise on (and in) the 3-D circle and accepts counter-clockwise radiation, (-c) CBR. The vertical line from the center to the detector has two representations. As a line connecting the four circles, it represents the world lines of the Earth and the elements that make up the detector. If the inside of the Z=0 circle had been cross-hatched, it would have represented the radius of the 4-D core at the detector. Therefore the telescope-detector always must point perpendicular to its radius to the center of the 4-D core.

For Fig. 4-3 and similar figures it will be assumed (A-2) that at emission the source of radiation was at rest ($v_{p=0}$) in the comoving frame of our universe. That assumption allows further relations to be shown.

Now consider an imaginary plane to have been cut through the 4-D core as fixed by the axis of the telescope and its radius to the center of the 4-D core. Past imprints on that slice through the 4-D core are what is shown in Fig. 4-3.

That slice passes through source 1 at Z=999 and the world line of that source stays on that slice as the universe expands and the (-c) CBR radiation imprints its spiral world line to our detector. All curves are computer generated and in particular the spiral for the radiation $v_c$=Hr-C follows H(t) from the SC-model R(t).

In agreement with Fig. 4-2, the emission distance ED measured on the circle labeled "Z=1.70", represents the maximum emission distance (at $Z_m$) and that relation would hold true if the detector, and its spiral curve, were moved around the universe to any other position on the "present" Z=0 circle.

To illustrate better both the reception distance RD and the emission distance ED, a second source 2 was placed in Fig. 4-3 on the Z=1.70 circle at the intersection with the spiral world line. This shows graphically that the ED arc distance to the vertical world line agrees with Figs. 4-1 and 4-2 and RD=(1+Z)ED. The 4-D analysis will now be expanded to obtain some extraordinary astronomical predictions from the SC-model.

## ASTRONOMICAL PREDICTIONS

If the universe is flat as present cosmologists prefer, then except for minor scattering and gravitational lensing, the world lines of radiation from a single source can never cross. On the other hand if our 3-D universe is closed as predicted by the SC-model, world lines of radiation from a single source can cross on the opposite side of the universe resulting in some extra-ordinary revelations and some spectacular events as will be described next.

**The Anoka-Scope** The source 1 radiation in Fig. 4-3 was emitted counter-clockwise but that source also emits radiation clockwise and future stars at that position could also emit clockwise. So what are the conditions where clockwise radiation from this source would also arrive simultaneously at the position of our detector? That condition is shown in Fig. 4-4.

The spiral world line for the clockwise radiation must be a mirror image through the vertical world line of the detector and it intersects the source 1 world line at Z=6.55 for this case of $t_0$=15 Gy ($R_0$=1.1x10$^{28}$ cm). But this radiation arrives at the detector from the opposite direction the detector is pointed.

Thus the CBR counter-clockwise radiation was emitted when our universe was only 1/(1+Z)=1/1000 times its present size and has traveled about 2/3 around our universe before reaching our detector at the present time. As our universe evolved and grew to 1/(1+Z)~1/8 its present size, stars were born in the same region of space and optical light was emitted in all directions including the future position of our detector. But this new radiation, in the direction opposite that of the previous CBR, had to travel only about 1/3 around our universe (against much smaller H) to arrive at our detector at the same time as the CBR radiation. For example, suppose the microwave image shows a large bright area surrounded by a ring of dark area and near it is a medium size totally dark area. In that case the optical image in the opposite direction might appear as a void surrounded with a shell of proto-galaxies and close by a medium size cluster of proto-galaxies.

When the author realized the importance of this finding, that two rays of radiation emitted in opposite directions from the same position in space and at different ages could reach our detectors at the same time, a new orbiting astronomical instrument was conceived and given the name of: "Anoka-Scope". At one end, a CBR-type instrument would be mounted to image the structure of the cosmic background microwave radiation and pointed in the exact opposite direction would be mounted an instrument for an optical image of the same solid angle of the same structures at redshift $Z_{same}$. In Sioux Indian, the name "anoka" means "on either side". The anoka-scope would be used to find that red shift $Z_{same}$ where both an original and evolved structure appeared the same "from either side". Once $Z_{same}$ is found, from the cosmological principle, the same result should be measured independent of the direction the instrument was pointed – excluding, of course, nearby objects that block a clear view of the cosmos like our own Galaxy.

Figure 4-4 was an early study before the SC-model was reduced to a single adjustable parameter, the age of the universe. So the study was repeated and Fig. 4-5 shows the predicted $Z_{same}$ over the age range of 12≤$t_0$, Gy≤ 16. with higher predicted values. The predicted value for $t_0$=13.5 Gy is $Z_{same}$≈9.0. As expected, $Z_{same}$ is very sensitive to the size of a closed universe and therefore to its age.

The current big bang model with its $\Omega_0$=1 flat universe cannot possibly produce a $Z_{same}$ as described above. On the other hand, a fixed $Z_{same}$ would not only be strong evidence that our universe is a 3-sphere, but would also fix its size and other cosmological parameters. With the new astronomical instruments under design and construction, astronomers may be able to make such measurements in the near future.

**Virtual Cosmic Background Structure** Are there other falsifiable predictions that can be gleaned from the 4-D representations of radiation trajectories in a closed but expanding universe? The answer is "Yes!" and the following predictions are even more extraordinary.

Figures 4-3 and 4-4 show the Z=999 radiation emitted in only one direction. Figure 4-6 shows the Z=999 radiation emitted in both directions in our expanding slice through the 4-D universe. The (1+Z)=1000 clockwise radiation ($v_c$=Hr+C) proceeds on to Z=0 far removed from our detector, but that is not the interesting event. The interesting event is the cross over point of the (+c) and (-c) spiral world lines on the opposite side of the universe from the source at Z~2. This fact becomes very exciting when one realizes that this cross over is not just true for two particular packets of radiation, one of which arrives at our detector, but the cross over point is the same for <u>every</u> packet of radiation from a single source that was emitted in all of the other directions at the same time as the two packets in this expanding slice of the 4-D universe. Thus the cross-over point becomes a "virtual source" of "apparent redshift" $Z_v$~2. The "real" redshift would be measured as $Z_e$=999, so the specious red shift $Z_v$ could be obtained only from the size of the refocused radiation and its disparate $Z_e$.

As another visual aid to accepting this important point, a very useful tool of a 3-D analogy will be used in a 3-D thought experiment where practicality is suppressed to make a theoretical point. All of the 24 meridian lines radiating away from the North Pole of the Earth terminate at its antipode, the South Pole. Now visualize 24 very fast airplanes in a circle above 89° North Latitude speeding away from the North Pole above the 24 meridians except the Earth is to expand to 333 times its size by the time they all reach the South pole at 89° South Latitude whose radius has now expanded to 333 times its original size (sufficient time for each plane to reach a different altitude for cross-over at the pole.)

So far, this analogy misses a very important point that can be corrected – replace each of the airplanes with a swarm of 24 airplanes. Instead of diverging from the North Pole, each swarm diverges from the point where its meridian (e.g., 90° W) crosses the 89° North Latitude circle. Now that swarm will not converge at the South Pole but on the circle at 89° South Latitude in the opposite hemisphere (e.g., 90° E). This addition to the analogy demonstrates that the planes diverging from the circle at 89° North Latitude all converge to the expanded circle at 89° South Latitude and then diverge from that circle. Also it shows that rotational parity is conserved – moving clockwise from A to B at the source, must be clockwise from A to B in the "virtual source".

In the 4-D case, source 1 is at the "North Pole" of a very small 4-D ball at the origin in Fig. 4-6 of radius 1/1000 the radius of the Z=0 ball or $r_e$=4.83 Mpc. All of the radiation is "speeding away" towards the cross over point ("South Pole") of radius $r_{co}$=4830/((1+Z)=3)=1610 Mpc. Only after being re-focused at the cross over "South Pole", do the radiation packets continue on their 4-D meridian paths and a small fraction of that now diverging radiation arrives at our detector many billion of years later.

Few people, if any, can clearly visualize four-dimensional objects, but it might aid acceptance of the 4-D claims by imagining rotation of the Z=0, 4-D ball of Fig. 4-6 around an axis consisting of the Source 1 world line extended to a diameter that included the source and the cross-over point. As our detector, and the spiral trajectory of its accepted radiation rotate out of the figure, all of the new radiation appearing from a circle around the source also generate the same two spiral-shaped world lines through the same cross-over point.

These predictions and conclusions have little to do with the basic SC-concept of spatial condensation. Instead they follow from the concept of a closed expanding universe. Only the specific value of $Z_{same}$ depends upon the R(t) of the SC-model.

Before stating further consequences, a definition of "front" and "back" of a distant structure needs to be defined. The (+c) spiral curve crossed the detector world line very early at about Z~9. We will call that CBR emission from the "front side" of whatever structure might have been there; of course, there could not have been any of our detectors there to measure it. Therefore the CBR radiation that we finally do measure at Z=0 comes through the virtual source from the "back side" of the source.

We have just concluded that the actual size of the of CBR structures at (1+Z)=1000 are 3/1000 the size we measure them today and are "backside" images of the last scattering surfaces. From the SC-expansion forces described in paper 3 [3], those images are certainly not gravitationally bound structures produced by very early quantum fluctuations. From the SC-model, those dark images are probably gravitationally unbound distributions of dark mass seeds that individually have accreted some matter.[3]

Now pause to consider the consequences of what has just been concluded. First remember that the CBR of Z=999 is of microwave frequency, far removed from the optical frequency, and all of its radiation with structure is being re-focused at Z≈2 relative to our detector. Different instruments are used to measure CBR and optical radiation. Thus stars and galaxies at Z≈2 do not appear at microwave frequency amongst the CBR structure, and conversely. Therefore the SC-model predicts the appearance of <u>virtual</u> CBR structure at $Z_v$~2 and magnified in size by a factor of $1000/(1+Z_v) \approx 333$. Next we search for even more spectacular virtual sources.

**Virtual Black Holes** In paper 3 [3], Figure 3-5 showed that dark mass could form black holes as early, or earlier, than (1+Z)=1000 and begin swallowing the hot, more dense, early plasma even before the decoupling of radiation and matter and may account for the quasars in active galactic nuclei (AGN) that we see today. Thus the spiral world line of z~1000 radiation could contain accretion disk radiation of not only late black holes but some as early as Z=1000 and most importantly, those accretion disks at very high redshift would appear to be those of "virtual black holes". The x-ray spectrum covers a frequency range of $10^{16}$ to $10^{20}$ Hz, so "hard" x-rays emitted at $10^{19}$ Hz could appear from virtual sources as $10^{16}$ Hz "soft" X-rays.

Black holes, particularly those in AGN at the center of galaxies, emit an enormous amount of energy over a wide range of the frequency spectrum. Unlike the CBR, they are small local emitters of energy. However because they are so bright, the radiation of very early black holes, reduced in frequency by a factor of $1/(1+Z_e)$, may also be measured by our detectors as "virtual black holes", expanded in size by the factor $(1+Z_e)/(1+Z_v)$ and reduced in flux intensity by the same factor.

Virtual black holes, measurable in their degraded x-ray spectrum, amongst real sources at Z~1 is one of the most extra-ordinary predictions of the SC-model. Measurements confirming such a prediction for our small, closed (Ω=0.3) universe expanding into a much larger epi-universe would be electrifying. Such a discovery could hardly be in greater contrast to the currently accepted big-bang theory of an infinite, flat (Ω=1) "spacetime" that is expanding – but denied any "space" (not even an empty one) to expand into.

There are geometric restrictions on which early black holes can be "seen" as virtual sources. From Fig. 4-6, the (+c) and (-c) spiral trajectories must cross and that can only happen if the actual source of radiation is a fraction f greater than or equal to one-half (f ≥0.5) around the universe from the detector. Although it appears very small on the scale of Fig. 4-6, the real black holes that form virtual sources must lie on the tiny segment of the (-c) spiral curve for a clockwise angle greater than 180 degrees from the detector. Real and virtual black holes might also be seen in the opposite direction on the mirror image of the (-c) spiral curve shown.

A downside to this prediction is to consider black holes that formed <u>on</u> the vertical world line of sources at f=0.5 whose virtual sources fall on our f=0 world line. A past example for an emission redshift $Z_e$~1000 source can be visualized by rotating both spiral curves together to bring the Z=2 crossover point on top of our world line. A more ominous example can be seen by rotating just the (+c) spiral curve counter-clockwise until the crossover point moves up and onto the Earth while the emitter moves to a $Z_e$<999 . As we will see, this happens when $Z_e$=34 and corresponds to a beam of radiation that could be devastating. Thus this scenario might be added to the list of possible causes for the demise of the dinosaurs. Virtual black holes could have crossed our past world line from age $t_0$=4.6 Gy to the present.

On the other side of the universe, all past black holes from about Ze=1000 to $Z_e$=34 could have had their virtual black holes wondering around our side of the 3-D universe. The odds are slim that one of these past black holes passed over the antipode of the Earth to make its virtual black hole cross the world line of the Earth. A rough estimate of the time interval for passing over of a direct hit can be obtained as follows: Assume a black hole with mass equal to one million times the mass of the Sun and an emitting accretion disk thirty times the Schwarzschild radius of diameter D≈$2 \times 10^8$ km. Further assume the virtual black hole diameter is amplified and its luminosity decreased by a factor of $(1+Z_v)/(1+Z_0)$=35. Also assume the black hole had a peculiar velocity sufficient to produce a virtual peculiar velocity of 200 km/s. This would produce a maximum lapse time of about one year for the refocused accretion disk to pass directly over the Earth and ignores the $1/r^2$ radiation during approach and retreat – terrible devastation even without a direct hit.

However, lets now return to more optimistic predictions for virtual black holes that do not cross our world line. If these virtual black holes can be seen, where do we look for them? What is the relation between the virtual $Z_v$ and the emission $Z_e$ for virtual black holes that are a considerable distance ($Z_v$) from us? The answer is presented in Fig. 4-7.

Figure 4-7 uses the abscissa for three types of curves. The lower curve labeled "$Z_v$" is the important curve of $Z_v$ versus $Z_e$ and uses the abscissa and the left ordinate. The curve above, for the same values of $Z_e$, shows the clockwise fraction f of the circumference of our universe, using the right ordinate, for the position of the real source. Again note that all virtual sources have f≥0.5. Both of these curves were developed for an SC-universe of age $t_0$=13.5 Gy.

The curve labeled (2) uses the left ordinate but the abscissa has units of age of the universe: and they are from top to bottom in units of Gy: 12,13,13.5,14,15 and 16. The real sources for these six points are all at (1+Z)=1000. Thus the third point down for $t_0$=13.5 Gy is the projection of curve (1) out to Z=999.

The striking feature of this Fig. 4-7 is that it places most of the virtual sources at low $Z_v$ values 0.5 – 1.5, in the middle of the Z-range of the galaxies observed today. So how do we tell the virtual from the real sources? Certainly, their spectrum redshift would be different (34<$Z_e$≤1000) but otherwise, the answer seems to be a difference in the size of structure. However, the reader is reminded that it takes only one such confirmed virtual source to precipitate a very major change in the present understanding of the evolution of our universe.

**Diffuse Extragalactic Light.** Real galaxies are difficult enough to "see" at $Z_e\sim6$, but what happens to the photons from stars and proto-galaxies at redshift $Z_e$=40 to 1000? There is a difference in their fate to us depending upon whether our 3-D universe is open or closed.

Galaxies emit radiation over a wide range of wavelength from radio waves to X-rays but we will be interested here in their emission in the infrared, visible and ultraviolet. Even more specific we are interested in detecting the amount of diffuse extragalactic light in the infrared K-band centered on $\lambda$=2190 nm. Because of redshift with distance, measurement of the K-band radiation can include sources from galaxies: (1) U-band at Z=5.00, (2) B-band at Z=3.92, (3) V-band at Z=2.97 and (4) R-band at Z=2.33 as well as I-, J-, H-, and K-band radiation from nearby galaxies. [11]

First consider the big-bang model where present cosmologists seem determined to force our universe to be a flat $\Omega$=1 universe. If a sufficiently large representative volume of this universe containing N galaxies, then one should be able to add all N galactic K-band contributions <u>in that volume V</u> and get the measured value of the K-band energy. This has been done carefully, assuming the flat model, and the answers do not agree – there is significantly more K-band (and J-band) radiation than the galaxy contributions. [12]

Before we go to the SC-model, there is a very important point to be noted for this flat model. Many of these N galaxies could emit appreciable ultraviolet radiation that leaves the volume V that eventually decays to K-band radiation. However, for the point to be made later, only such incoming radiation should be counted if it enters the narrow solid angle of the detector and that should be negligible from such a distant source.

Now consider a similar volume V for the closed SC-model. The conclusion about the ultraviolet radiation emitted outside volume V is altogether different. The distant ultraviolet emission is the one particular wavelength, produced in abundance by the proto-galaxies at the antipode of volume V, that will travel one half way around our closed 3-D universe and <u>all</u> of the proto-galaxy's ultraviolet radiation will produce virtual sources of K-band (and J-band) radiation in our volume V.

The expanded virtual sources will be too weak to be recognized as galaxies but their diffuse radiation will be there to contribute to the total and should account for some of the excess measured.

## A PREFERRED REFERENCE FRAME

The reader who has followed the development of this new spatial condensation cosmological model in the last three papers will have recognized very little of the basic concepts of relativity theory even though many predictions are in agreement. Agreements with the big-bang progeny of relativity theory include: the scaling of radiation and matter with the expansion, our 3-D universe begins with omega equal to unity, the new time and the density of matter must correctly predict the era of nucleosynthesis of the light elements, radiation and matter must decouple at Z~1000 to release the CBR and the new source of gravity predicts the same curvature of 3-D space as the hyperspace of relativity theory.

Disagreements with relativity theory began at the very beginning of the new model with the exclusion of gravity as having anything to do with either the creation or expansion of our universe. With our 3-D universe as the surface of an expanding 4-D ball, our universe could not possibly collapse as predicted by general relativity. A fourth spatial dimension is completely foreign to relativity theory as is the production of cells of space that can be counted and cannot just expand.

In the discussion of the transport of radiation above, the reader also may have noticed an underlying assumption of privilege given to the comoving frame of reference where the speed of light is C. The prediction of the relativists that a massive object with a peculiar velocity $v_p$ is destined to come to rest in the comoving frame [5,7] was accepted and built into the expansion force of paper 3 [3] where the drag force per unit mass or acceleration $a_H = -(1+C_t)Hv_p$

It follows that a fundamental question now is: "<u>In principle</u>, can one, from within an *inertial frame* (e.g., rocket ship + instrument, power off) measure one's peculiar velocity and then bring the rocket ship to rest in the preferred frame?" If so, this would violate Einstein's special theory of relativity.

This question was analyzed and answered in the affirmative elsewhere.[8,9] The means to find one's peculiar velocity and to come to rest in the comoving frame is shown in the embedded sketch in Fig. 4-8 and this thought experiment need only establish the principle and not the practicality of such an experiment.

The rocket ship is equipped with extendable rods, ostensibly rigid fore and aft, of equal length, with a mirror on the far end of each. The ship also contains the necessary instruments to send simultaneously a pulse of radiation into the vacuum (rocket power off) towards each of the mirrors at $X=\pm X_m$ and also two receptors to record the reflected signals and any difference in the two round-trip times-of-flight. This one-dimensional experiment is meant to determine one component, say in the X-direction, of its absolute velocity in the comoving frame. After adjusting the x-velocity to $v_x=0$, the pilot re-orients the rocket ship to an orthogonal Y-direction, brings $v_y=0$ and the repeats in the Z-direction to finally come to rest in the comoving frame.

Consider what is happening in this thought experiment. From Paper 3 [3] remember that both 4-D core pk and 3-D pk spatial cells are being produced (vacuum energy without mass) in every volume of our 3-D space and it is also this production that accounts for expansion of our universe and Hubble's law. Also remember that it is this outflow of 3-D space from any position in 3-D space that creates a drag on any massive object that is prevented from participating in this outward flow.

In the sketch of Fig.4.8 consider first that the center of the spaceship is at rest in this expanding 3-D space with its extended rods at their equal maximum lengths. Space is being produced in and all around the space ship and its extended rods and is flowing in all directions from its point sources. But note that position $X=+X_c$ ahead of the spaceship, 3-D space is flowing in and around the front rod mass at $+X_c$ at velocity $v(X_c)=+HX_c$. Equivalently, even though at rest with the center of the spaceship, the rod mass can be said to be moving at peculiar velocity $v_p(+X_c)=-HX_c$ relative to 3-D space. Also on the trailing rod, $v(-X_c)=-HX_c$ and $v_p(-X_c)=+HX_c$.

Next suppose the center of the spaceship is drifting forward at a peculiar velocity in the comoving frame, that is, the center is moving through the stationary space at velocity $v_{p0}=+HX_c$. In that case the forward velocity of the rod mass at $X=X_c$ cancels its previous negative peculiar velocity and now has come to rest in the comoving frame. On the other hand, the rod mass at $X=-X_c$ and has increased in peculiar to $v_p(-X_c)=+2HX_c$.

This change in the velocity of the spaceship should certainly change the times-of-flight of two new light signals which we now try to calculate but we should not be surprised that the predicted change is very small.

The analysis needs only Eq. (1), expressed in terms of this thought experiment. However, caution is needed in this exercise as the role of the peculiar velocity is considered. From our vantage point of the comoving frame for this analysis, the pilot has a peculiar velocity called $v_x$. For a forward velocity $v_x$, the net peculiar velocity, $v_{px}+v_x=0$ at $X_C=v_x/H$. So the compounded velocity of light at position X is,

$$v_c=+H(X-X_C)\pm C. \tag{4}$$

The +C is used for the emitted pulse and –C is used for the reflected signal. The time lapse from $X_1$ to $X_2$ is,

$$\Delta t=\int dX/[XH+(C-HX_C)], \tag{5}$$

from $X_1$ to $X_2$. The round-trip solutions $\Delta t(-m)$ and $\Delta t(+m)$ [8,9] give the difference in round-trip times-of-flight,

$$Dt=\Delta t(-m)-\Delta t(+m), \tag{6}$$

and using the approximation $\ln(1\pm X)=\pm X$ and defining $\gamma=1/(1-(v_x/C)^2)^{1/2}$ gives the approximate round-trip times-of-flight,

$$Dta = 4\gamma^4 H(X_m/C)^2(v_x/C), \tag{7}$$

and Dta goes to zero in the limits of $v_x=X_m=H=0$ as it should.

The results of the calculation for the difference in round-trip times-of-flight are presented in the lower curve of Fig. 4-8 where the symbols are calculations for Dt and the curve is the approximation Dta. In order to obtain reasonable differences in round-trip times-of-flight of order seconds, the distance to mirrors $L_m$ in Fig. 4-8 was set to the un-reasonably large distance of $L_m=1$ pc. Nevertheless for reasonable $L_m$, the effect is predicted to be there in principle even though we cannot measure it. The upper curve surprisingly shows that the difference between the forward and backward times-of-flight is about five orders of magnitude greater than the differences in round-trip times-of-flight.

The Earth at distance of 1 Au, orbits the Sun at about 30 km s$^{-1}$ ($v/C\sim 1\times 10^{-4}$) and the Earth-Mars distance varies from about 0.5 to 1.5 Au. Thus if astronauts could take one of their good atomic clocks to Mars and get it synchronized with one on Earth, they might be able to check the predicted difference in radar times-of-flight of about 0.11 seconds [9] for the condition of a null Mars-Earth relative velocity.

The author thanks his good friend, Emeritus Professor Robert A. Piccirelli, for extensive discussions of the new physical concepts.

**FUTURE PAPERS**

Many of the basic concepts and predictions of the spatial condensation model in the last four papers are not consistent with the current theory of relativity. However these good and reasonable predictions of the SC-model are announcing that the relativistic spacetime model does not embrace enough spatial dimensions and its basic concepts of space, time and energy have no deeper understanding and each of these basic concepts must itself be modeled to forge the desired compatibility with quantum theory. The author fully understands that the physics community is very reluctant to modify its present theory and that more than a good competing theory may be necessary to stimulate a paradigm change.

The SC-model is still in its formative stage particularly with respect to its unification with quantum behavior. Future papers may follow as the model develops.

## REFERENCES

1. Leffert, C. B. 2001, http://xxx.lanl.gov/abs/astro-ph/0102071
2. Leffert, C. B. 2001, http://xxx.lanl.gov/abs/astro-ph/0102318
3. Leffert, C. B. 2001, http://xxx.lanl.gov/abs/astro-ph/0106236
4. Zinkernagel, H. 2001, http://xxx.lanl.gov/abs/astro-ph/0105130
5. Kolb, E. W., & Turner, M. S. 1990, *The Early Universe* (Reading: Addison-Wesley) p 38, p84.
6. Harrison, E. R. 2000, *Cosmology* (Cambridge: Cambridge University Press) p319, 447.
7. Peebles, P. J. E. 1993, *Principles of Physical Cosmology* (Princeton: Princeton Univ. Press) p 95.
8. Leffert, C. B. 1995, *Time and Cosmology: Creation and Expansion of Our Universe* (Troy: Anoka Publishing), ISBN 0-9647745-6-9.
9. Leffert, C. B. 1999 *Evolution of Our Universe: via Spatial Condensation* (Troy: Anoka Publishing) ISBN 0-9647745-9-3.
10. Leffert, C. B. & Donahue, T. M. 1958, Am. J. Phy., **26**, 515.
11. Binney, J. & Merrifield, M. 1998, *Galactic Astronomy* (Princeton: Princeton Univ. Press) p57.
12. Totani, T., et. al., 2001, http://xxx.lanl.gov/abs/astro-ph/0102328

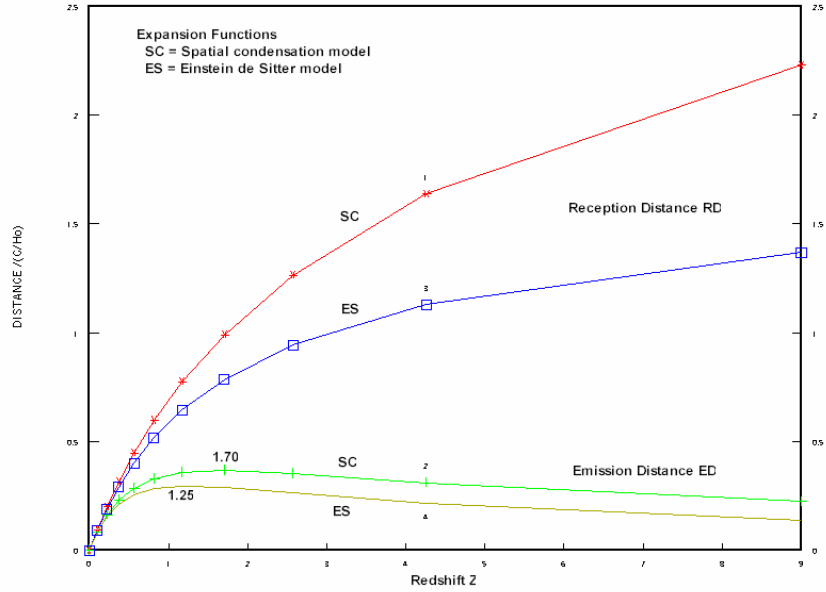

Fig. 4-1. Expansion of the universe increases the distance to any past source of received radiation and decreases its energy (increases its wave length redshift Z). As a function of Z, the predicted emission distance ED from past sources to our world line, goes through a maximum and it, and the present reception distance RD to those sources, depend upon the model of the universe R(t) as shown for the Einstein de Sitter ES and the new spatial-condensation SC models.

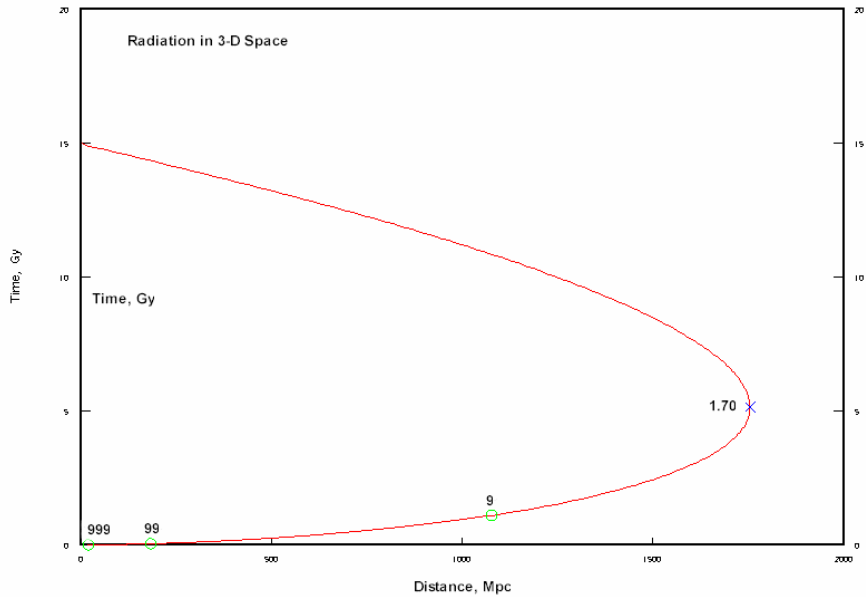

Fig. 4-2. A predicted 3-D space-time, SC-example trajectory is shown for a packet of cosmic background radiation CBR emitted at decoupling, (1+Z)=1000, with present arrival at 15 Gy. Even though emitted only ~67 light years (ly) from our world line, the rapid expansion carried it away to 5.72 billion ly at Z=1.70 before the radiation could finally decrease the distance to our detector.

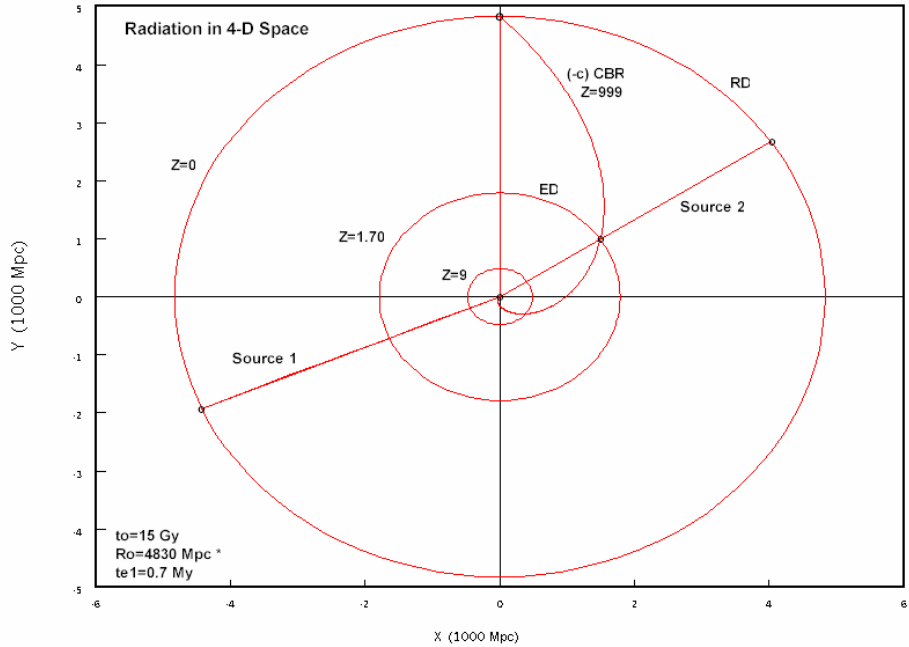

Fig. 4-3. This spiral curve in 4-D space shows the same SC-example trajectory as in Fig. 4-2 of a packet of radiation (emitted counter-clockwise) where the four circles represent only four slices through the expanding 4-D core. The arc distance labeled ED is indeed the maximum arc distance suffered by the radiation. All radiation entering the detector (pointed clockwise on the Z=0 meridian) must lie on the spiral radiation world line.

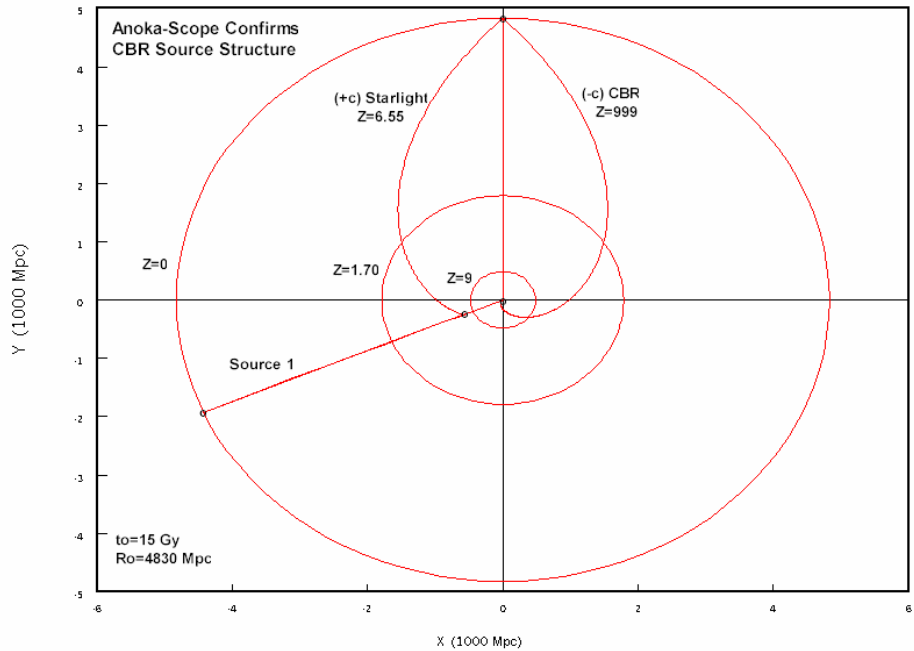

Fig. 4-4. For the same example of Fig. 4-3, the CBR source 1 evolves into a galaxy and the starlight emitted clockwise at $Z_{same}$=6.55 arrives simultaneously with the CBR radiation (from the opposite direction). Large-scale images are compared in a new orbiting "Anoka-Scope" instrument to measure $Z_{same}$ independent of its alignment in 3-D space to fix the size of the closed expanding universe.

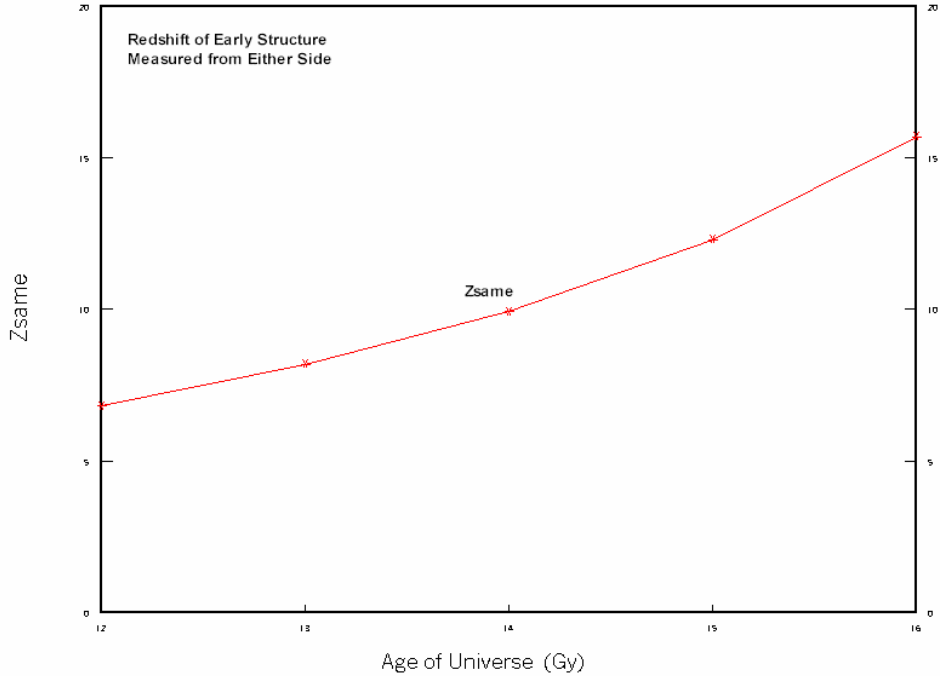

Fig. 4-5. The predicted redshift $Z_{same}$ of Fig. 4-4 is presented for the advanced SC-model versus the one adjustable parameter, the age of the universe, $t_0$, Gy.

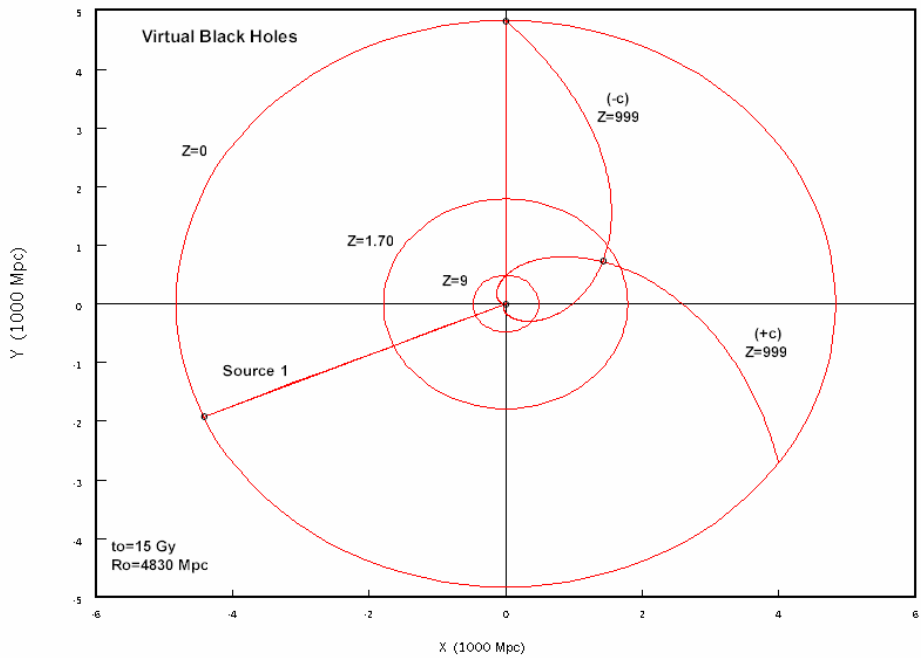

Fig. 4-6. Predicted dark-mass black holes [2] at decoupling, $(1+Z)=1000$, would produce virtual black holes as indicated by the cross over of the (+c) and (-c) spiral curves. For this example, the radiation $Z=999$ and distance $Z_v \sim 2$ should be measurable since all (free) black hole radiation should focus at the cross over magnified in size by the factor $(1+Z)/(1+Z_v)=333$.

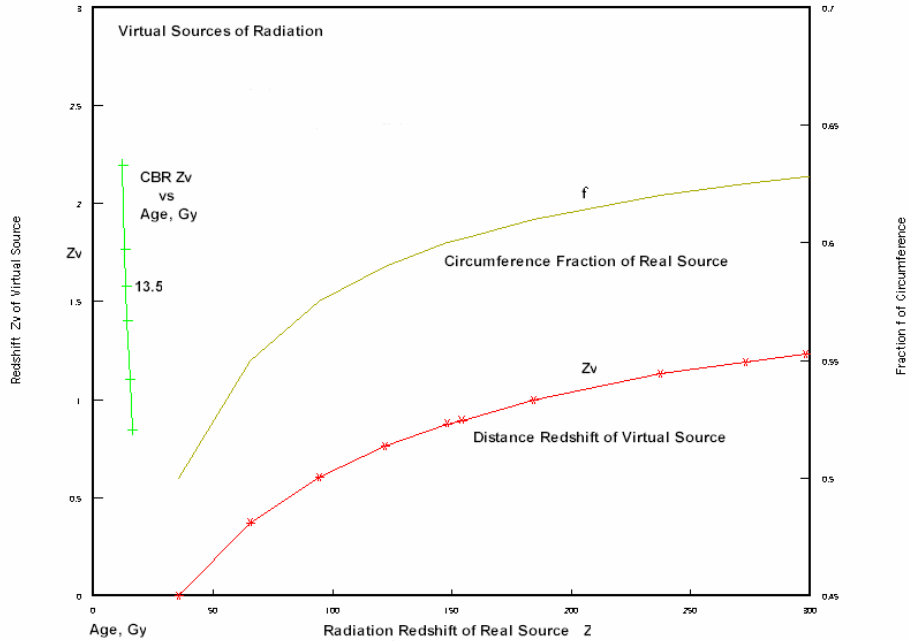

Fig. 4-7. The relation of the distance $Z_v$ of virtual black holes on the (-c) spiral curve of Fig. 4-6 to radiation redshift Z of the real black hole source is shown by the curve labeled "$Z_v$" for a $t_0$=13.5 Gy SC-universe. The clockwise position from the source from the detector is at fraction "f" of the circumference. Note the limited predicted range $0.5 \leq Z_v \leq 1.5$ of virtual black holes among most of the visible galaxies

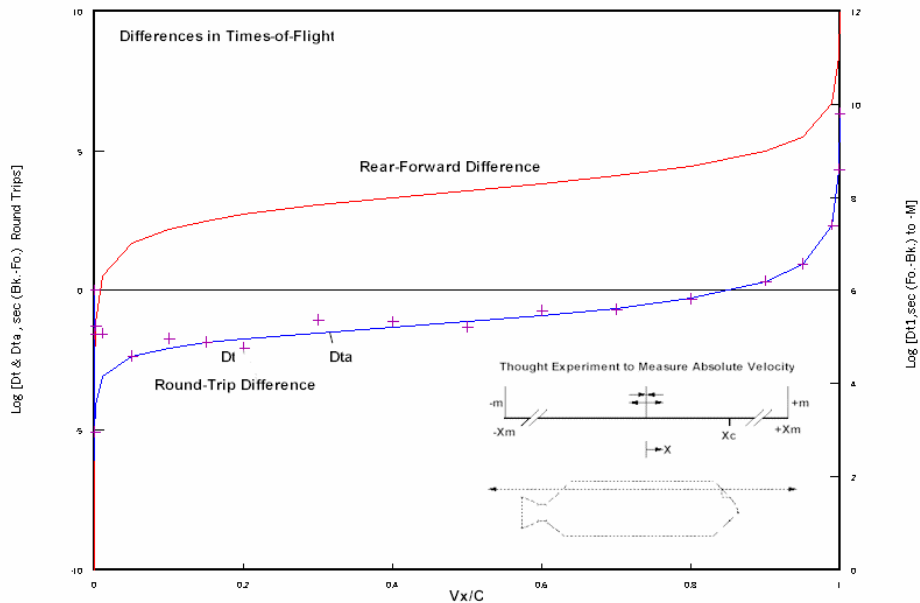

Fig. 4.8. A reasonable difference of seconds in round-trip times-of-flight Dt at reasonable peculiar velocities of $v_x/C<0.1$, required unreasonable extendable rod lengths of $X_m \sim 1$ pc. Nevertheless, the principle is demonstrated of finding the preferred inertial frame. Surprisingly, the rear-forward difference in flight times was some 5 orders of magnitude larger and suggests a future experimental check (see text).